\documentstyle[times,pramana,epsf,floats]{ias}
\newcommand{\NP}[1]{ Nucl.\ Phys.\ {#1}}

\newcommand{\PR}[1]{Phys.\ Rev.\ {#1}}

\newcommand{\La}{{\cal L}}

\newcommand{\be}{\begin{equation}}
\newcommand{\ee}{\end{equation}}
\newcommand{\ba}{\begin{eqnarray}}
\newcommand{\ea}{\end{eqnarray}}

\begin{document}

\mark{{}{}}
\title{Chiral dynamics of baryon resonances and hadrons in a nuclear medium}

\author{E. Oset, D. Cabrera, V.K. Magas, L. Roca, S.Sarkar, M.J. Vicente Vacas
and A. Ramos$^1$ }
\address{Departamento de F\'{\i}sica Te\'orica and IFIC,
Centro Mixto Universidad de Valencia-CSIC,
Institutos de Investigaci\'on de Paterna, Apdo. correos 2085,\\
46071, Valencia, Spain.\\
$^1$ Departament d'Estructura i 
Constituents de la Mat\'eria, \\
Universitat de Barcelona,
 Diagonal 647, 08028 Barcelona, Spain}
\abstract{ In these lectures I make an introduction to chiral unitary theory
applied to the meson baryon interaction and show how several well known
 resonances are dynamically generated, and others are predicted. Two very recent
 experiments are analyzed, one of them showing the existence of two
 $\Lambda(1405)$ states and the other one providing support for the 
 $\Lambda(1520)$ resonance
 as a quasibound state of $\Sigma(1385) \pi$. The use of 
 chiral  Lagrangians to account for the  hadronic
 interaction at the elementary level introduces a new approach to deal with the
 modification of meson and baryon properties in a nuclear medium.  Examples of
 it for $\bar{K}$, $\eta$ and $\phi$ modification in the nuclear medium are
 presented.   
}

\maketitle
\section{Introduction}
Nowadays it is commonly accepted that QCD is the theory of the strong
interactions, with the quarks as building blocks for baryons and mesons, and
the gluons as the mediators of the interaction. However, at low energies typical
of the nuclear phenomena, perturbative calculations with the QCD Lagrangian are
not possible and one has to resort to other techniques to use the information of
the QCD Lagrangian. One of the most fruitful approaches has been the use of
chiral perturbation theory, $\chi PT$ \cite{xpt}. The theory introduces effective
Lagrangians which involve only observable particles, mesons and baryons,
respects the basic symmetries of the original QCD Lagrangian, particularly
chiral symmetry, and organizes these effective Lagrangians according to the
number of derivatives of the meson and baryon fields.  

The introduction of unitarity constraints in coupled channels in chiral
perturbation theory has led to unitary extensions of the theory that starting
from the same effective Lagrangians allow one to make predictions at much higher
energies . One of the interesting
consequences of these extensions is that they generate dynamically low lying
resonances, both in the mesonic and baryonic sectors. By this we mean that they
are generated by the multiple scattering of the meson or baryon components,
much  the same as the deuteron is generated by the interaction of the nucleons
through the action of a  potential, and they are not preexistent states that
remain in the large $N_c$ limit where the multiple scattering is suppressed. 

\section{Baryon meson interaction}\label{baryon}
  The interaction of the octet of stable baryons with the octet of pseudoscalar
  mesons is given to lowest order by the Lagrangian \cite{ulf,ecker}
\begin{eqnarray}
\label{BaryonL}
\La_1=<\bar{B}i\gamma^\mu \bigtriangledown_\mu B >-M_B
<\bar{B}B>&+&\frac{1}{2}D<\bar{B}\gamma^\mu \gamma_5
\left\{u_\mu,B \right\}>\\
&+&\frac{1}{2}F<\bar{B}\gamma^\mu \gamma_5 [u_\mu,B] >\nonumber
\label{eq:lowest}
\end{eqnarray}  

\begin{equation}
\begin{array}{l}
\nabla_{\mu} B = \partial_{\mu} B + [\Gamma_{\mu}, B] \\
\Gamma_{\mu} = \frac{1}{2} (u^+ \partial_{\mu} u + u \partial_{\mu} u^+) \\
U = u^2 = {\rm exp} (i \sqrt{2} \Phi / f) \\
u_{\mu} = i u ^+ \partial_{\mu} U u^+
\end{array}
\end{equation}
where the symbol $<>$ stands for the trace of the matrices and 
the SU(3) matrices for the meson and the baryon fields are given by

\begin{equation}
\Phi =
\left(
\begin{array}{ccc}
\frac{1}{\sqrt{2}} \pi^0 + \frac{1}{\sqrt{6}} \eta & \pi^+ & K^+ \\
\pi^- & - \frac{1}{\sqrt{2}} \pi^0 + \frac{1}{\sqrt{6}} \eta & K^0 \\
K^- & \bar{K}^0 & - \frac{2}{\sqrt{6}} \eta
\end{array}
\right)
\end{equation}

\begin{equation}
B =
\left(
\begin{array}{ccc}
\frac{1}{\sqrt{2}} \Sigma^0 + \frac{1}{\sqrt{6}} \Lambda &
\Sigma^+ & p \\
\Sigma^- & - \frac{1}{\sqrt{2}} \Sigma^0 + \frac{1}{\sqrt{6}} \Lambda & n \\
\Xi^- & \Xi^0 & - \frac{2}{\sqrt{6}} \Lambda
\end{array}
\right)
\end{equation}

At lowest order in momentum, that we will keep in our study, the interaction
Lagrangian comes from the $\Gamma_{\mu}$ term in the covariant derivative
and we find

\begin{equation}
L_1^{(B)} = < \bar{B} i \gamma^{\mu} \frac{1}{4 f^2}
[(\Phi \partial_{\mu} \Phi - \partial_{\mu} \Phi \Phi) B
- B (\Phi \partial_{\mu} \Phi - \partial_{\mu} \Phi \Phi)] >
\end{equation}

\noindent
which leads to a common structure of the type
$\bar{u} \gamma^u (k_{\mu} + k'_{\mu}) u$ for the different channels, where
$u, \bar{u}$ are the Dirac spinors and $k, k'$ the momenta of the incoming
and outgoing mesons.

We take the $K^- p$ state and all those that couple to it within the chiral
scheme. These states are $\bar{K}^0 n, \pi^0 \Lambda, \pi^0 \Sigma^0,
\pi^+ \Sigma^-, \pi^- \Sigma^+, \eta \Lambda, \eta \Sigma^0, K^+ \Xi^-, K^0
\Xi^0$. Hence we have
a problem with ten  coupled channels. We should notice that, in addition to
the six channels considered in \cite{Kaiser:1995eg} one has the two $\eta$ channels,
$\eta \Lambda$ and $\eta \Sigma^0$ and the two $K$ channels, 
$K^+ \Xi^-, K^0 \Xi^0$. Although these channels are above
threshold for $K^- p$ scattering at low energies, they couple strongly to the
$K^- p$ system and there are important interferences between the real parts
of the amplitudes, which make their inclusion in the coupled-channel approach
very important.

The lowest order amplitudes for these channels are easily evaluated from eq.
(5) and are given by

\begin{equation}
V_{i j} = - C_{i j} \frac{1}{4 f^2} \bar{u} (p') \gamma^{\mu} u (p)
(k_{\mu} + k'_{\mu})
\end{equation}

\noindent
were $p, p' (k, k')$ are the initial, final momenta of the baryons (mesons).
Also, for low energies one can safely neglect the spatial components in eq.
(6) and only the $\gamma^0$ component becomes relevant, hence simplifying
eq. (6) which becomes

\begin{equation}
V_{i j} = - C_{i j} \frac{1}{4 f^2} (k^0 + k'^0)
\label{poten}
\end{equation}

with  $C_{i j}$  SU(3) coefficients, which are easily derived from the chiral
Lagrangians  \cite{kaon}.

 \section{Unitarized chiral perturbation theory: N/D or dispersion relation
method}

  One can find a systematic and easily comprehensible derivation 
 of the  ideas of the N/D method applied for the first time to the meson baryon system in
 \cite{Oller:2000fj}, which we reproduce here below and which follows closely
 the similar developments used before in the meson meson interaction \cite{nsd}.
 One defines the transition $T-$matrix as $T_{i,j}$ between the coupled channels which couple to
 certain quantum numbers. For instance in the case of  $\bar{K} N$ scattering studied in
 \cite{Oller:2000fj} the channels with zero charge are $K^- p$, $\bar{K^0} n$, $\pi^0 \Sigma^0$,$\pi^+
 \Sigma^-$, $\pi^- \Sigma^+$, $\pi^0 \Lambda$, $\eta \Lambda$, $\eta \Sigma^0$, 
 $K^+ \Xi^-$, $K^0 \Xi^0$.
 Unitarity in coupled channels is written as
 
\begin{equation} 
Im T_{i,j} = T_{i,l} \rho_l T^*_{l,j}
\end{equation}
where $\rho_i \equiv 2M_l q_i/(8\pi W)$, with $q_i$  the modulus of the c.m. 
three--momentum, and the subscripts $i$ and $j$ refer to the physical channels. 
 This equation is most efficiently written in terms of the inverse amplitude as
\begin{equation}
\label{uni}
\hbox{Im}~T^{-1}(W)_{ij}=-\rho(W)_i \delta_{ij}~,
\end{equation}
The unitarity relation in Eq. (\ref{uni}) gives rise to a cut in the
$T$--matrix of partial wave amplitudes, which is usually called the unitarity or right--hand 
cut. Hence one can write down a dispersion relation for $T^{-1}(W)$ 
\begin{equation}
\label{dis}
T^{-1}(W)_{ij}=-\delta_{ij}\left\{\widetilde{a}_i(s_0)+ 
\frac{s-s_0}{\pi}\int_{s_{i}}^\infty ds' 
\frac{\rho(s')_i}{(s'-s)(s'-s_0)}\right\}+{\mathcal{T}}^{-1}(W)_{ij} ~,
\end{equation}
where $s_i$ is the value of the $s$ variable at the threshold of channel $i$ and 
${\mathcal{T}}^{-1}(W)_{ij}$ indicates other contributions coming from local and 
pole terms, as well as crossed channel dynamics but {\it without} 
right--hand cut. These extra terms
are taken directly from $\chi PT$ 
after requiring the {\em matching} of the general result to the $\chi PT$ expressions. 
Notice also that 
\begin{equation}
\label{g}
g(s)_i=\widetilde{a}_i(s_0)+ \frac{s-s_0}{\pi}\int_{s_{i}}^\infty ds' 
\frac{\rho(s')_i}{(s'-s)(s'-s_0)}
\end{equation}
is the familiar scalar loop integral.

One can further simplify the notation by employing a matrix formalism. 
Introducing the 
matrices $g(s)={\rm diag}~(g(s)_i)$, $T$ and ${\mathcal{T}}$, the latter defined in 
terms 
of the matrix elements $T_{ij}$ and ${\mathcal{T}}_{ij}$, the $T$-matrix can be written as:
\begin{equation}
\label{t}
T(W)=\left[I-{\mathcal{T}}(W)\cdot g(s) \right]^{-1}\cdot {\mathcal{T}}(W)~.
\end{equation}
which can be recast in a more familiar form as 
 \begin{equation}
\label{ta}
T(W)={\mathcal{T}}(W)+{\mathcal{T}}(W) g(s) T(W)
\end{equation}
Now imagine one is taking the lowest order chiral amplitude for the kernel
${\mathcal{T}}$ as done in
\cite{Oller:2000fj}. Then the former equation is nothing but the Bethe Salpeter equation with the
kernel taken from the lowest order Lagrangian and  factorized  on  shell, the same
approach followed in \cite{kaon}, where different arguments were used to justify the on shell
factorization of the kernel.

The on shell factorization of the kernel, justified here with the N/D method,
renders the set of coupled Bethe Salpeter integral equations a simple set of
algebraic equations.

 \section{Meson baryon scattering}
The low-energy $K^-N$ scattering and transition to coupled channels is one of
the cases of successful application of chiral dynamics in the baryon
sector.  We rewrite Eq. (7) in the more familiar form 
\begin{equation}
T = V + V \, G \, T
\label{eq:bs2}
\end{equation}
with $G$ the diagonal matrix given by the loop function of a meson and a baryon
propagators.

The analytical expression for $G_l$ can be obtained from \cite{kaon} using a
 cut off and from \cite{Oller:2000fj} using dimensional regularization. One has
 
\be
G_{l} = i \, \int \frac{d^4 q}{(2 \pi)^4} \, \frac{M_l}{E_l
(\vec{q})} \,
\frac{1}{\sqrt{s} - q^0 - E_l (\vec{q}) + i \epsilon} \,
\frac{1}{q^2 - m^2_l + i \epsilon}
\label{gfn}
\ee
in which $M_l$ and $m_l$ are the masses of the baryons and mesons respectively.
In the dimensional regularization scheme this is given by
\begin{eqnarray}
G_{l}&=& i \, 2 M_l \int \frac{d^4 q}{(2 \pi)^4} \,
\frac{1}{(P-q)^2 - M_l^2 + i \epsilon} \, \frac{1}{q^2 - m^2_l + i
\epsilon}  \nonumber \\ 
&=& \frac{2 M_l}{16 \pi^2} \left\{ a_l(\mu) + \ln
\frac{M_l^2}{\mu^2} + \frac{m_l^2-M_l^2 + s}{2s} \ln \frac{m_l^2}{M_l^2}
-2i\pi \frac{q_l}{\sqrt{s}}
\right. \nonumber \\ & &  \phantom{\frac{2 M}{16 \pi^2}} +
\frac{q_l}{\sqrt{s}}
\left[
\ln(s-(M_l^2-m_l^2)+2 q_l\sqrt{s})+
\ln(s+(M_l^2-m_l^2)+2 q_l\sqrt{s}) \right. \nonumber  \\
& & \left. \phantom{\frac{2 M}{16 \pi^2} +
\frac{q_l}{\sqrt{s}}}
\left. \hspace*{-0.3cm}- \ln(s-(M_l^2-m_l^2)-2 q_l\sqrt{s})-
\ln(s+(M_l^2-m_l^2)-2 q_l\sqrt{s}) \right]
\right\},
\label{propdr}
\end{eqnarray}
where $\mu$ is the scale of dimensional regularization, $a_l$ is the
subtraction constant and $q_l$ denotes the 
three-momentum of the meson or baryon in the centre of mass frame.
We then look for poles of the transition matrix $T$ in the complex $\sqrt{s}$ 
plane. The complex poles, $z_R$, appear in unphysical Riemann sheets.
 We use the loop function of Eq. (\ref{propdr}) for energies below
threshold and replace it above threshold with
\be
G_l^{2nd}=G_l+2i\,\frac{q_l}{\sqrt{s}}\,\frac{M_l}{4\pi}
\label{defR2}
\ee
where the variables on the right hand side of the above equation are
evaluated in the first (physical) Riemann sheet.
It is trivially verified that the same can be achieved by 
changing the sign of the complex valued momentum $q_l$ from positive 
to negative in the loop function $ G_l(z)$ of
eq.~(\ref{propdr}) for the channels which
are above threshold at an energy equal to Re($z$).
This we call the second
Riemann sheet $R_2$.

\section{Strangeness $S= -1$ sector}

We take the $K^- p$ state and all those that couple to it within the chiral
scheme mentioned above.  Hence we have a problem with ten coupled channels.
The coupled set of Bethe Salpeter equations 
were solved in \cite{kaon} using a cut off momentum of 630
MeV in all channels. Changes in the cut off can be accommodated in terms
 of changes in $\mu$, the regularization scale in the dimensional
 regularization formula for  $G_l$, or in the subtraction constant
$a_l$. In
 order to obtain the same results as in \cite{kaon} at low energies, we set
 $\mu$ equal to the cut off momentum of 630 MeV (in all channels) and then
find the values of the
 subtraction constants $a_l$ such as to have $G_l$ with the same value
with the
 dimensional regularization formula  and the cut
off formula at the $\bar{K} N$ threshold. 
For the purpose of this talk let us recall that in \cite{kaon} we 
obtain the $\Lambda(1405)$ resonance 
obtained from the $\pi \Sigma$ spectrum and the cross sections for 
$K¯p $ to different channels, some of which are shown in fig. 1.

\begin{figure}[htb]
\vspace*{-0.5 cm}
\epsfxsize=12cm
\centerline{\epsfbox{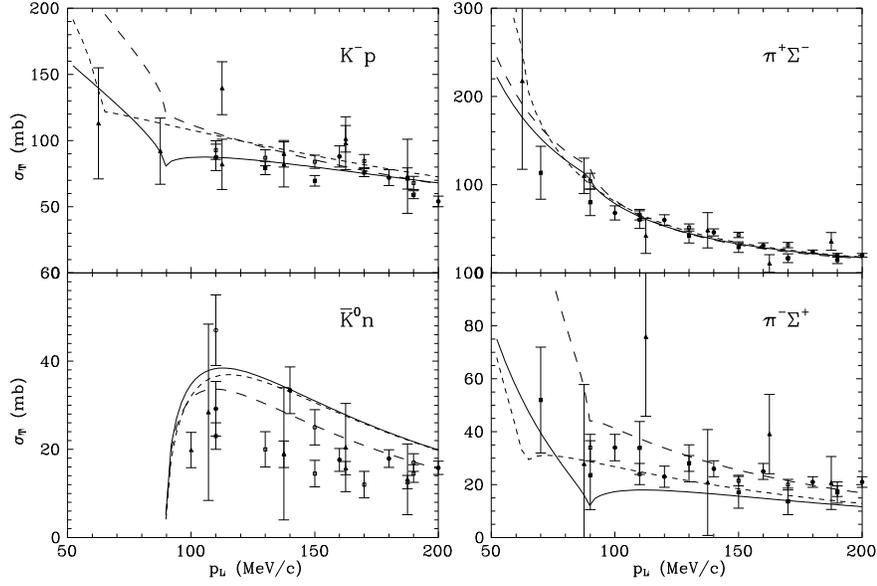}}
\caption{
$K^-p$ scattering cross sections as functions of the $K^-$
momentum in the lab frame: with the full basis of physical states
(solid line), omitting the $\eta$ channels (long-dashed line) and
with the isospin-basis (short-dashed line). Taken from
Ref.~\protect\cite{kaon}.
\label{fig:kncross}}
\end{figure}

\section{Strangeness $S= 0$ sector}

  The strangeness $S= 0$ channel was also investigated using the Lippmann
Schwinger equation and coupled channels in \cite{Kaiser:1995cy}. 
The $N^*(1535)$ resonance was also generated dynamically within this 
approach. Subsequently work was done in this sector in \cite{Nacher:1999vg},
and \cite{Nieves:2001wt} where the $N^*(1535)$ 
resonance was also generated. 
In \cite{Inoue:2001ip} the work along these lines was continued and improved
 by introducing the $\pi N \to \pi NN$ channels, which proved essential in
 reproducing the isospin 3/2 part of the $\pi N$ amplitude, including the
 reproduction of the $\Delta(1620)$ resonance.

  For total zero charge one has six channels in this case,
 $\pi^- p$, $\pi^0 n$, $\eta n$, $K^+ \Sigma^-$,
 $K^0 \Sigma^0$, and $K^0 \Lambda$.   
 
 Details on the issues discussed in this
 paper can be seen in the review paper \cite{review}.
 
 \section{Poles of the T-matrix}
\label{sec:3}

  The study of Ref.~\cite{bennhold} showed the presence of  poles in
  Eq.~(\ref{t}) around the $\Lambda(1405)$ and the
$\Lambda(1670)$ for isospin $I=0$ and around the $\Sigma(1620)$ in
$I=1$.  The same approach in $S=-2$ leads to the resonance
$\Xi(1620)$ \cite{xi} and in $S=0$ to the $N^*(1535)$
\cite{Kaiser:1995cy,Inoue:2001ip}. 
 One is thus tempted to consider the appearance of a singlet and an octet
of meson--baryon resonances. Nevertheless, the situation is more
complicated because indeed in the SU(3) limit there are {\it two} octets
and not just one, as we discuss below. 
The presence of these multiplets was already
  discussed in Ref.~\cite{Oller:2000fj} after obtaining a pole with $S=-1$ in the
$I=1$ channel,
with mass around 1430 MeV, and two poles with $I=0$, of masses around that of the
  $\Lambda(1405)$.

 The appearance of a multiplet of dynamically generated mesons and baryons seems
 most natural once a state of the multiplet appears. Indeed, one must recall that the
 chiral Lagrangians are obtained from the combination of the octet of
 pseudoscalar mesons (the pions and partners) and the octet of stable baryons
(the nucleons and partners).  The SU(3) decomposition of the combination of two
 octets tells us that
 \begin{equation}
 8 \otimes 8=1\oplus 8_s \oplus 8_a \oplus 10 \oplus \overline{10} \oplus 27~.
\end{equation}
Thus, on pure SU(3) grounds, should we have a SU(3) symmetric Lagrangian,
 one can expect e.g. one singlet and two octets of resonances, the symmetric and
 antisymmetric ones.  

 The lowest order of the meson--baryon chiral Lagrangian is exactly SU(3) 
invariant if
 all the masses of the mesons, or equivalently the quark
 masses,  are set equal.  
In Ref.~\cite{bennhold}
 the baryon masses take their physical values, although strictly
 speaking at the leading order in the chiral expansion they should be equal to
 $M_0$. For Eq.~(\ref{poten}) being SU(3) symmetric, all the baryons masses
 $M_{i}$ must be set equal as well. When all the meson and baryon masses are
  equal, and these common masses are employed in evaluating the $G_l$ functions,
 together with equal subtraction constants $a_l$, the $T$--matrix obtained
from Eq.~(\ref{t}) is also SU(3) symmetric. 

If we do such an SU(3) symmetry approximation
and look for poles of the scattering matrix, we find poles
corresponding to the octets and singlet. The surprising result is that
the two octet poles are degenerate as a consequence of the 
 dynamics contained in
 the chiral Lagrangians. Indeed, if we evaluate the matrix elements of the transition potential
$V$ in a basis of SU(3) states, we obtain something proportional to 
$V_{\alpha  \beta}= {\rm diag}(6,3,3,0,0,-2)$
taking the following order for the irreducible representations:
$1$, $8_s$, $8_a$, $10$, $\overline{10}$ and $27$, with positive sign meaning
attraction.

Hence we observe that the states belonging to different
irreducible representations do not mix and the two octets appear
degenerate. The coefficients in $V_{\alpha  \beta}$
 clearly illustrate why there are no bound
states in the $10$, $\overline{10}$ and $27$ representations.

In practice, the same chiral Lagrangians allow for SU(3) breaking. In
the case of Refs.~\cite{kaon,bennhold} the breaking appears
because both in the $V_{i j}$ transition potentials as in the $G_l$
loop functions one uses the
physical masses of the particles as well as different subtraction constants in $G_l$, 
corresponding to the use of a unique cut-off in all channels. 
 In Ref.~\cite{Oller:2000fj} the
physical masses are also used in the $G_l$ functions, although these functions are evaluated 
with a unique subtraction constant as corresponds to the SU(3) limit. 
In both approaches, physical masses are
used to evaluate the $G_l$ loop functions so that unitarity is fulfilled
exactly and the physical thresholds for all channels are respected. 

\begin{figure}
\epsfxsize=12cm
\centerline{\epsfbox{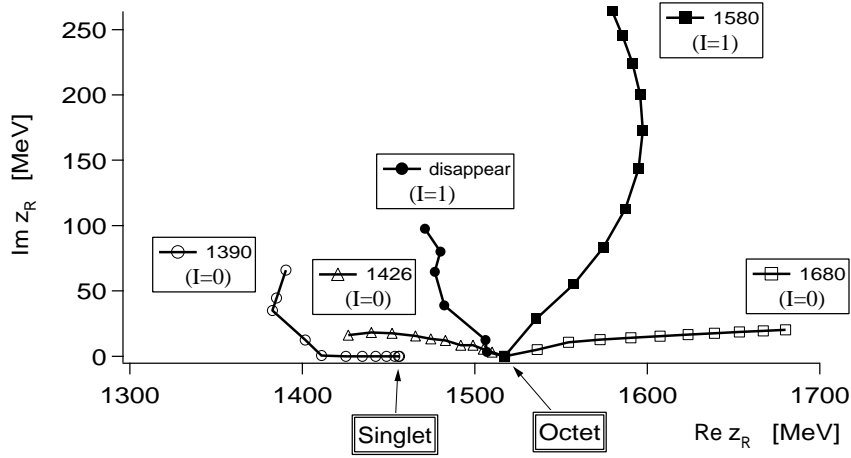}}
  \caption{Trajectories of the poles in the scattering amplitudes obtained by
  changing the SU(3) breaking parameter $x$ gradually. At the SU(3) symmetric 
  limit ($x=0$),
   only two poles appear, one is for the singlet and the other for the octet.
  The symbols correspond to the step size $\delta x =0.1$. The results are from
  \protect\cite{Jido:2003cb}.}
  \label{fig:tracepole}
\end{figure}

By following the approach of Ref.~\cite{bennhold} and using the
physical masses of the baryons and the mesons, the position of the
poles change and the two octets split apart in four branches, two
for $I=0$ and two for $I=1$, as one can see in \cite{Jido:2003cb}, which we
reproduce in 
Fig.~\ref{fig:tracepole}. In the figure we show the trajectories
of the poles as a function of a parameter $x$ that breaks
gradually the SU(3) symmetry up to the physical values.  The
dependence of masses and subtraction constants on the parameter
$x$ is given by
\begin{eqnarray}
M_i(x) &= & M_0+x(M_i-M_0),  \nonumber \\
m^{2}_{i}(x) &=& m_{0}^{2} + x (m^{2}_{i}-m^{2}_{0}), \nonumber\\
a_{i}(x) &=& a_{0} + x (a_{i} - a_{0}),
\end{eqnarray}
where $0\le x \le 1$.  In the calculation of
Fig.~\ref{fig:tracepole}, the values $M_{0}=1151$ MeV, $m_{0} =
368$ MeV and $a_{0}= -2.148 $ are used.

  The complex poles, $z_R$, appear in unphysical sheets. In the
present search we follow the strategy of changing
the sign of the momentum $q_l$ 
in the $G_l(z)$ loop function  for the channels which
are open at an energy equal to Re($z$).

The splitting of the two $I=0$ octet states is very interesting.
One moves to higher energies to merge with the $\Lambda(1670)$
resonance and the other one moves to lower energies to create a
pole, quite well identified below the  $\bar{K}N$ threshold, with
a narrow width. 
 On the other hand, the singlet also evolves
to produce a pole at low energies with a quite large width.

We note that the singlet and the $I=0$ octet states appear
nearby in energy and  what experiments  actually
see is a combination of the effect of these two resonances.

Similarly as for the $I=0$ octet states, we can see that one branch of the 
$I=1$ states moves to higher energies while another
moves to lower energies. The branch moving to higher energies finishes at
what would correspond to the $\Sigma(1620)$ resonance when the physical
masses are reached. The
branch moving to lower energies fades away after a while when getting close to
the $\bar{K}N$ threshold.  

The model of Ref.~\cite{Oller:2000fj} reproduces qualitatively the same results, but
the $I=1$ pole also stays for $x=1$. Nevertheless,
 in both approaches there
is an $I=1$ amplitude with an enhanced strength around the $\bar{K} N$ threshold.
 This amplitude has non negligible  consequences for 
reactions producing $\pi \Sigma$ pairs in that region.
This has been illustrated for instance
in Ref.~\cite{nacher1}, where the photoproduction of the $\Lambda(1405)$ via the
reaction $\gamma p \to K^+ \Lambda(1405)$ was studied. It was shown there that
the different sign in the
$I=1$ component of the $\mid \pi^+ \Sigma^-\rangle$, $\mid \pi^- \Sigma^+\rangle$
states leads, through interference between the $I=1$ and the dominant $I=0$
amplitudes, to different
cross sections in the various charge channels, a fact that has been
confirmed experimentally very recently \cite{ahn}.

Once the pole positions are found, one can also determine the
couplings of these resonances to the physical states by studying
the amplitudes close to the pole and identifying them with
\begin{equation}
T_{i j}=\frac{g_i g_j}{z-z_R}~.
\end{equation}
The couplings $g_i$ are in general complex valued numbers.
In  Table 1 we summarize the
pole positions and the complex couplings $g_i$ obtained from the
model of Ref.~\cite{bennhold} for isospin $I=0$. The results with the model of
\cite{Oller:2000fj} are qualitatively similar.

\begin{table}[ht]
\centering 
\caption{\small Pole positions and couplings to $I=0$
physical states from the model of Ref.~\protect\cite{bennhold}}
 \vspace{0.5cm}
\begin{tabular}{|c|cc|cc|cc|}
\hline
 $z_{R}$ & \multicolumn{2}{c|}{$1390 + 66i$} &
\multicolumn{2}{c|}{$1426 + 16i$} &
 \multicolumn{2}{c|}{$1680 + 20i$}  \\
 $(I=0)$ & $g_i$ & $|g_i|$ & $g_i$ & $|g_i|$ & $g_i$ & $|g_i|$ \\
 \hline
 $\pi \Sigma$ & $-2.5-1.5i$ & 2.9 & $0.42-1.4i$ & 1.5 & $-0.003-0.27i$ &
 0.27 \\
 ${\bar K} N$ & $1.2+1.7i$ & 2.1 & $-2.5+0.94i$ & 2.7 & $0.30+0.71i$ &
 0.77 \\
 $\eta\Lambda$ & $0.010+0.77i$ & 0.77 & $-1.4+0.21i$ & 1.4 & $-1.1-0.12i$ &
 1.1 \\
 $K\Xi$ & $-0.45-0.41i$ & 0.61 & $0.11-0.33i$ & 0.35 & $3.4+0.14i$ &
 3.5 \\
 \hline
 \end{tabular}
\label{tab:jido0}
\end{table}

 We observe that the
second resonance with $I=0$ couples strongly to $\bar{K} N$ channel, while
the first resonance couples more strongly to $\pi \Sigma$.

\section{Influence of the poles on the physical observables}
\label{sec:5}

In a given reaction the $\Lambda(1405)$ resonance is always seen in $\pi \Sigma$
mass distribution. However, the $\Lambda(1405)$ can be produced through any of
the channels in  Table 1. Hence,
it is clear that,  should there be a reaction which forces this
initial channel to be $\bar{K}N$, then this would give more
weight to the second resonance, $R_{2}$, and hence produce a
distribution with a shape corresponding to an effective resonance
narrower than the nominal one and at higher energy. Such a case
indeed occurs in the reaction $K^- p \to \Lambda(1405) \gamma$
studied theoretically in Ref.~\cite{nacher}.  It was shown there
that since the $K^- p$ system has a larger energy than the
resonance, one has to lose energy emitting a photon prior to the
creation of the resonance and this is effectively done by the
Bremsstrahlung from the original $K^-$ or the proton.  Hence the
resonance is initiated from the  $K^- p$ channel. This is also the case
in the reaction $\gamma p\to K^* \Lambda(1405)$ which has also the 
$\Lambda(1405)$ initiated by $\bar{K} N$ through the vertex 
$\gamma \to K^* K$ \cite{Hyodo:2004vt}. In the next section we report on a
recent experiment and its theoretical analysis that gives strong support to the
idea of the two $\Lambda(1405)$ states. 

\section{Evidence for the two pole structure of the 
$\Lambda(1405)$ resonance}

The recently measured reaction
$K^- p \to \pi^0 \pi^0 \Sigma^0$  \cite{Prakhov} allows us to test already the
two-pole nature of the $\Lambda(1405)$. This
process
shows a strong similarity with the reaction $K^- p \to \gamma \Lambda(1405)$, where the
photon is replaced by a $\pi^0$. 

\begin{center}
\begin{figure}[htb]
\epsfxsize=6cm
\centerline{\epsfbox{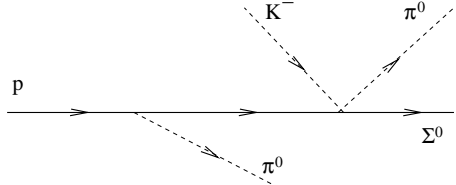}}
 \caption{
Nucleon pole term for the $K^- p \to \pi^0 \pi^0 \Sigma$ reaction.}
\label{fig:tree}
\end{figure}
\end{center}

Our model for the reaction 
$K^- p \to \pi^0 \pi^0 \Sigma^0 $
in the energy region of $p_{K^-}=514$ to $750$ MeV/c, as in the experiment \cite{Prakhov}, 
considers those mechanisms in which a $\pi^0$ loses the necessary energy
to allow the remaining $\pi^0\Sigma^0$ pair to be on top of the $\Lambda(1405)$
resonance.  The first of such mechanisms is given by the diagram of
Fig.~\ref{fig:tree}.
 In addition, analogy with the $\pi^- p \to K^0 \pi \Sigma$ reaction, where
the $\pi\Sigma$ is also produced in the $\Lambda(1405)$ region, demands that one
also considers the mechanisms with a meson pole, involving the meson meson 
amplitudes, as done in \cite{hyodo}.  These mechanisms, together with the one of fig. \ref{fig:tree}, are
considered in the work of \cite{Magas:2005vu}, which we summarize here, where the
meson pole term was found to give a very small contribution.

\begin{figure*}[htb]
\epsfxsize=15cm
\centerline{\epsfbox{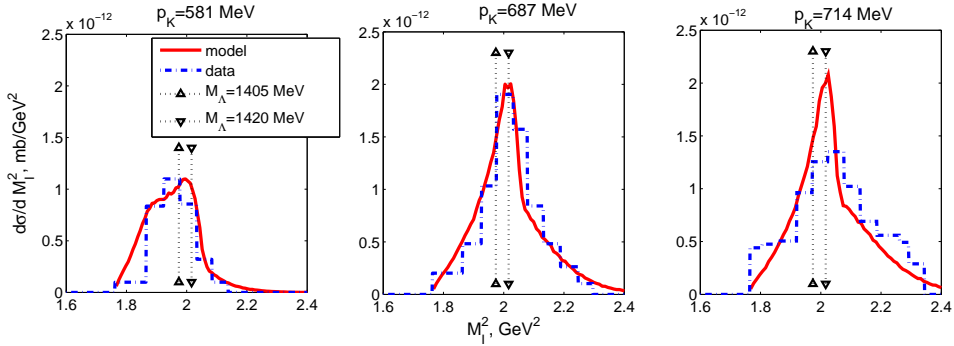}}
\caption{The ($\pi^0 \Sigma^0$) invariant mass distribution for three different initial kaon
momenta.}
 \label{fig:mass}
\end{figure*}

Our calculations show that the process is largely dominated by the nucleon
pole term shown in Fig.~\ref{fig:tree}. As a consequence, the $\Lambda(1405)$ thus obtained comes
mainly from the $K^- p \to \pi^0 \Sigma^0$ amplitude which, as mentioned above,
gives  the largest possible weight  to the second (narrower) state.


\begin{figure}[htb]
\epsfxsize=9cm
\centerline{\epsfbox{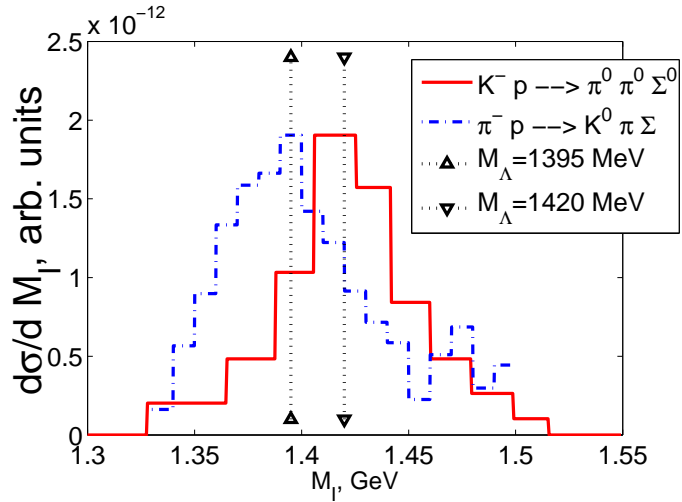}}
 \caption{Two experimental shapes of  $\Lambda(1405)$ resonance. 
 See text for more details. 
 \label{two_exp}
 }
\vspace{-0.5cm}
\end{figure}


In Fig.~\ref{fig:mass} our results for the invariant mass distribution
for three different energies of the incoming $K^-$ are
compared to the experimental data. Symmetrization of the amplitudes produces a
sizable amount of background. At a kaon laboratory momentum of $p_K=581$ MeV/c
this background  distorts the $\Lambda(1405)$ shape producing cross section in
the lower part of $M_I$, while at $p_K=714$ MeV/c the strength of this
background is shifted toward the higher $M_I$ region. An ideal situation is
found for momenta around $687$ MeV/c, where the background sits below the
$\Lambda(1405)$ peak distorting its shape minimally. The peak of the resonance
shows up at $M_I^2=2.02$ GeV$^2$ which corresponds to $M_I=1420$ MeV, larger
than the nominal $\Lambda(1405)$, and in agreement with the predictions of
Ref.~\cite{Jido:2003cb} for the location of the peak when the process is
dominated by the $t_{{\bar K}N \to \pi\Sigma}$ amplitude.  The apparent width
from experiment is about $40-45$ MeV, but a precise determination would require
to remove the background mostly coming from the ``wrong'' $\pi^0 \Sigma^0$
couples due to the indistinguishability of the two pions. A theoretical analysis
permits extracting the pure resonant part by not symmetrizing the amplitude,
and this is done in \cite{Magas:2005vu}, where it is found that 
the width of the resonant part is $\Gamma=38$ MeV, which
is smaller than the nominal $\Lambda(1405)$ width of $50\pm 2$ MeV \cite{PDG},
obtained from the average of several experiments, and much narrower than the
apparent width of about $60$ MeV that one sees in the $\pi^- p \to K^0 \pi
\Sigma$ experiment \cite{Thomas}, which also produces a distribution peaked at
$1395$ MeV.
In order to illustrate the difference between the $\Lambda(1405)$ resonance
seen in this latter reaction and in the present one, the two
experimental distributions are compared in Fig. \ref{two_exp}. We recall
that the shape of the  $\Lambda(1405)$ in the $\pi^- p \to K^0 \pi \Sigma$ 
reaction was shown in Ref.~\cite{hyodo} to be largely built from the
 $\pi \Sigma \to \pi \Sigma$ amplitude, which is dominated by
the wider, lower energy state.


\begin{figure}[htb]
\epsfxsize=8cm
\centerline{\epsfbox{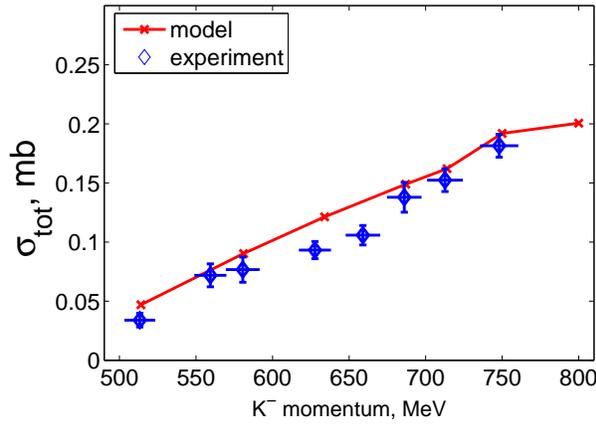}}
\caption{Total cross section for the reaction $K^- p \to
 \pi^0 \pi^0 \Sigma^0$. Experimental data are taken from
 Ref.~\protect\cite{Prakhov}.
 \label{fig:cross}
 }
\end{figure}


The invariant mass distributions shown here are not normalized, as in
experiment. But we can also compare our absolute values of
the total cross sections with those in Ref.~\cite{Prakhov}. As shown in 
Fig.~\ref{fig:cross}, our results are in excellent agreement with the data, 
in particular for the three kaon momentum values whose corresponding
invariant mass distributions have been displayed in
Fig.~\ref{fig:mass}.

In summary, we have shown, by means of a realistic model, that the  $K^- p \to
 \pi^0 \pi^0 \Sigma^0$ reaction is particularly suited to study the features
 of the second pole of the $\Lambda(1405)$ resonance, since it is largely
 dominated by a mechanism in which a $\pi^0$ is emitted prior to the $K^- p \to
 \pi^0 \Sigma^0$ amplitude, which is the one giving the largest weight to the
 second narrower state at higher energy.  
 Our model has proved to be accurate in reproducing both the
 invariant mass distributions and  integrated cross sections seen in 
 a recent experiment \cite{Prakhov}.  The study of the present
 reaction, complemental to the one of Ref.  \cite {hyodo} for the $\pi^- p \to
 K^0 \pi \Sigma$ reaction, has shown that the quite different shapes of the 
 $\Lambda(1405)$ resonance seen in these experiments can be interpreted in favour
 of the existence of two poles with the corresponding states having the
 characteristics predicted by the chiral theoretical calculations.  
 Besides demonstrating once more the great predictive power of the chiral
 unitary theories, this combined study of the two reactions gives the first
 clear evidence of the two-pole nature of the $\Lambda(1405)$.

\section{The interaction of the decuplet of baryons with the octet of mesons}
Given the success of the chiral unitary approach in generating dynamically low
energy resonances from the interaction of the octets of stable baryons
and the pseudoscalar mesons,   in 
\cite{Kolomeitsev:2003kt} the
interaction of the decuplet of $3/2^+$ with the octet of pseudoscalar mesons 
was studied and shown to
lead to many states which were associated to experimentally well 
established $3/2^-$ resonances. 
  
\begin{figure}[htbp]
\epsfxsize=15cm
\centerline{\epsfbox{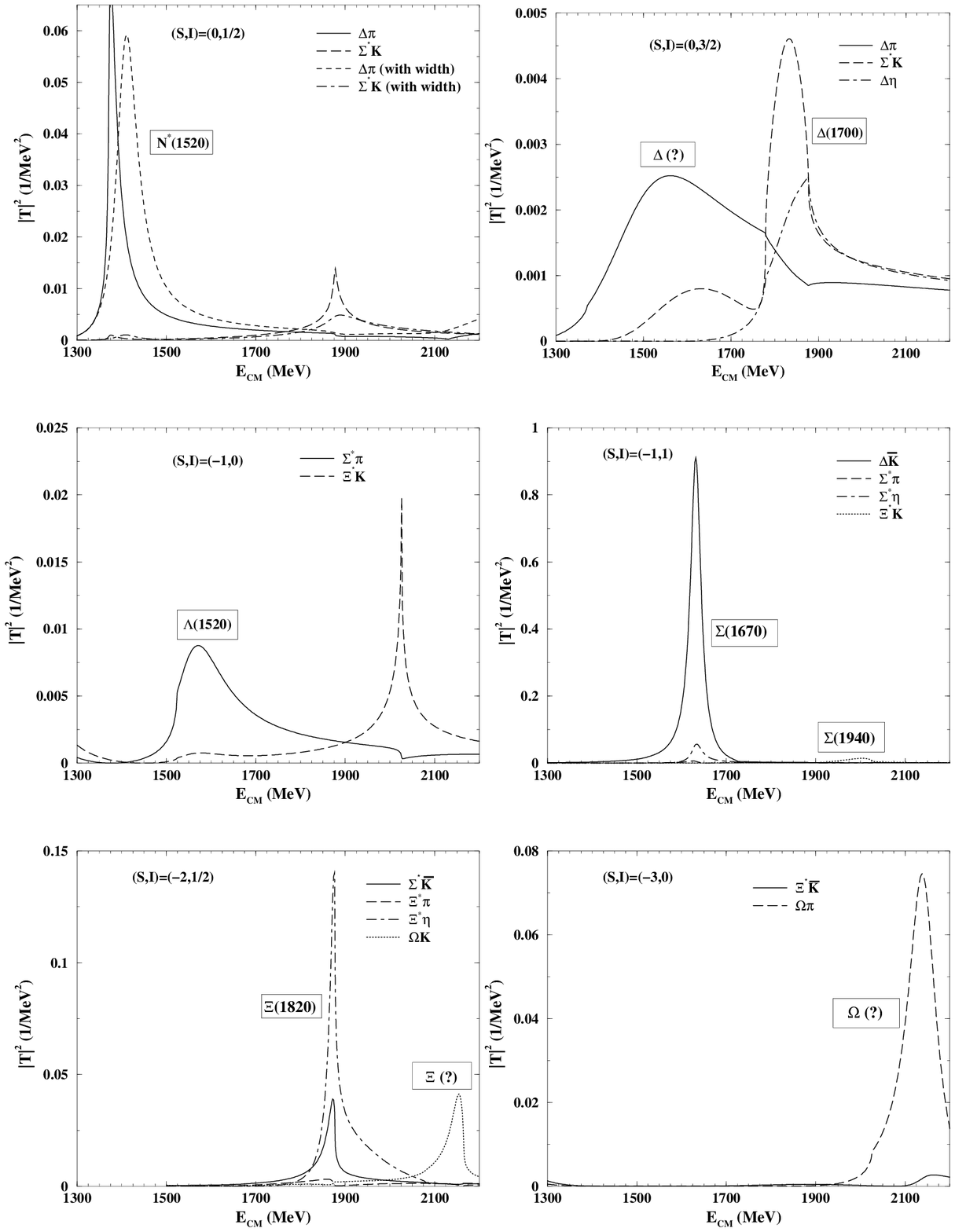}}
\caption{Resonances obtained from the interaction of the octet of mesons with
  the decuplet of baryons.}
  \label{fig:decu}
\end{figure}

    The lowest order chiral Lagrangian for the interaction of the baryon 
decuplet with the octet of pseudoscalar mesons is given by \cite{Jenkins:1991es}
\begin{equation}
L=-(i\bar T^\mu D_{\nu} \gamma^{\nu} T_\mu -m_T\bar T^\mu T_\mu)
\label{lag1} 
\end{equation}
where $T^\mu_{abc}$ is the spin decuplet field and $D^{\nu}$ the covariant derivative
given by in \cite{Jenkins:1991es}. The identification of the physical decuplet
states with the $T^\mu_{abc}$ can be seen in \cite{sarkar}, where a detailed
study of this interaction and the resonances generated can be seen. The study is
done along the lines of the former sections, looking for poles in the second
Riemann sheet of the complex plane, the coupling of the resonances to the
different channels and the stability of the results with respect to variations
of the input parameter, which in our case is just the subtraction constant, a.
This allows the association of the resonances found to existing states of the
particle data book, and the prediction of new ones.  A detail of the results
obtained can be seen in Fig. \ref{fig:decu}.

  Another interesting result is the generation of an exotic state of $S=1$ 
  and $I=1$ which  is generated  by the interaction of the $\Delta K $ channels
  and stands as a $\Delta K$ resonance. The pole appears in a Riemann sheet
  below threshold when also the sign of the momentum is changed, but it leads
  to a $\Delta K $ amplitude which accumulates strength close to threshold and 
produces a broad peak in the cross section  in contrast to the
$I=2$ cross section which is much smaller and very smooth.  However, the
situations is not completely clear since as shown in \cite{Sarkar:2004sc} the results
are very sensitive to changes in the input and the mere change of $f$ to $f_K$ 
makes the pole disappear.  It is not clear what would happen if higher order
terms of the interaction were taken into account, and for the moment it remains
as an interesting observation.

\section{Chiral coupled channel dynamics of the $\Lambda(1520)$ and the 
$K^-p\to \pi^0\pi^0 \Lambda$ reaction}

\begin{figure}[htbp]
\epsfxsize=5cm
\centerline{\epsfbox{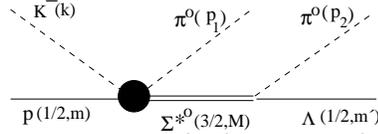}}
\caption{Scheme for $K^-p\to\pi^0\Sigma^{*0}(1385)\to\pi^0\pi^0\Lambda(1116)$. The
blob indicates the unitarized vertex.}
\label{kpfig}
\end{figure}

\begin{figure}[htbp]
\epsfxsize=9cm
\centerline{\epsfbox{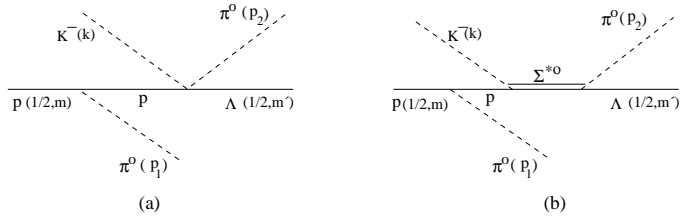}}
\caption{A conventional scheme for $K^-p\to\pi^0\pi^0\Lambda$}
\label{conv1}
\end{figure}

\begin{figure}[htbp]
\epsfxsize=8cm
\centerline{\epsfbox{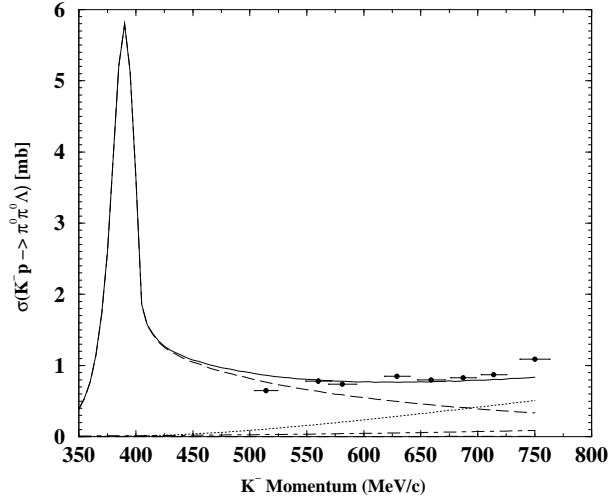}}
\caption{Cross-section as a function of the $K^-$ momentum. The dot-dashed 
and dotted lines are
the contributions of the diagrams of figs.~\ref{conv1}(a) and \ref{conv1}(b)
respectively. The
dashed line shows the cross section with fig.~\ref{kpfig} only 
and the solid line for a coherent sum of all these diagrams.}
\label{sigfig2}
\end{figure}

In \cite{Sarkar:2005ap} a refinement of the approach discussed above has been done
including the $\bar{K} N$ and $\pi \Sigma$ decay channels of the $\Lambda(1520)$ and
fine  tunning the subtraction constant in the g function, such that a good
agreement with the position and width of the $\Lambda(1520)$ is attained.  With this
new information one can face the study of the reaction $K^-p \to
\pi^0\pi^0\Lambda$ by
using the mechanisms of fig. \ref{kpfig} which provides the dominant contribution to
the reaction at energies close to the $\Lambda(1520)$.  At higher energies of the
experiment of \cite{Prakhov:2004ri} one finds that the mechanisms depicted in 
fig. \ref{conv1} provide
a contribution that helps bring the theory and experiment in good agreement as
we can see in fig. \ref{sigfig2}.  One can see in the figure that up to 575 MeV/c of
momentum of the $K^-$ the mechanism based on the strong coupling of the  
$\Lambda(1520)$ to the $\pi \Sigma(1385)$ channel is largely dominant and provides
the right strength of the cross section.  Obviously it would be most interesting
to investigate the region of lower energies of the $K^-$ in order to see if
the predictions done by the theory are accurate. 
 
\begin{figure}[htbp]
\epsfxsize=8cm
\centerline{\epsfbox{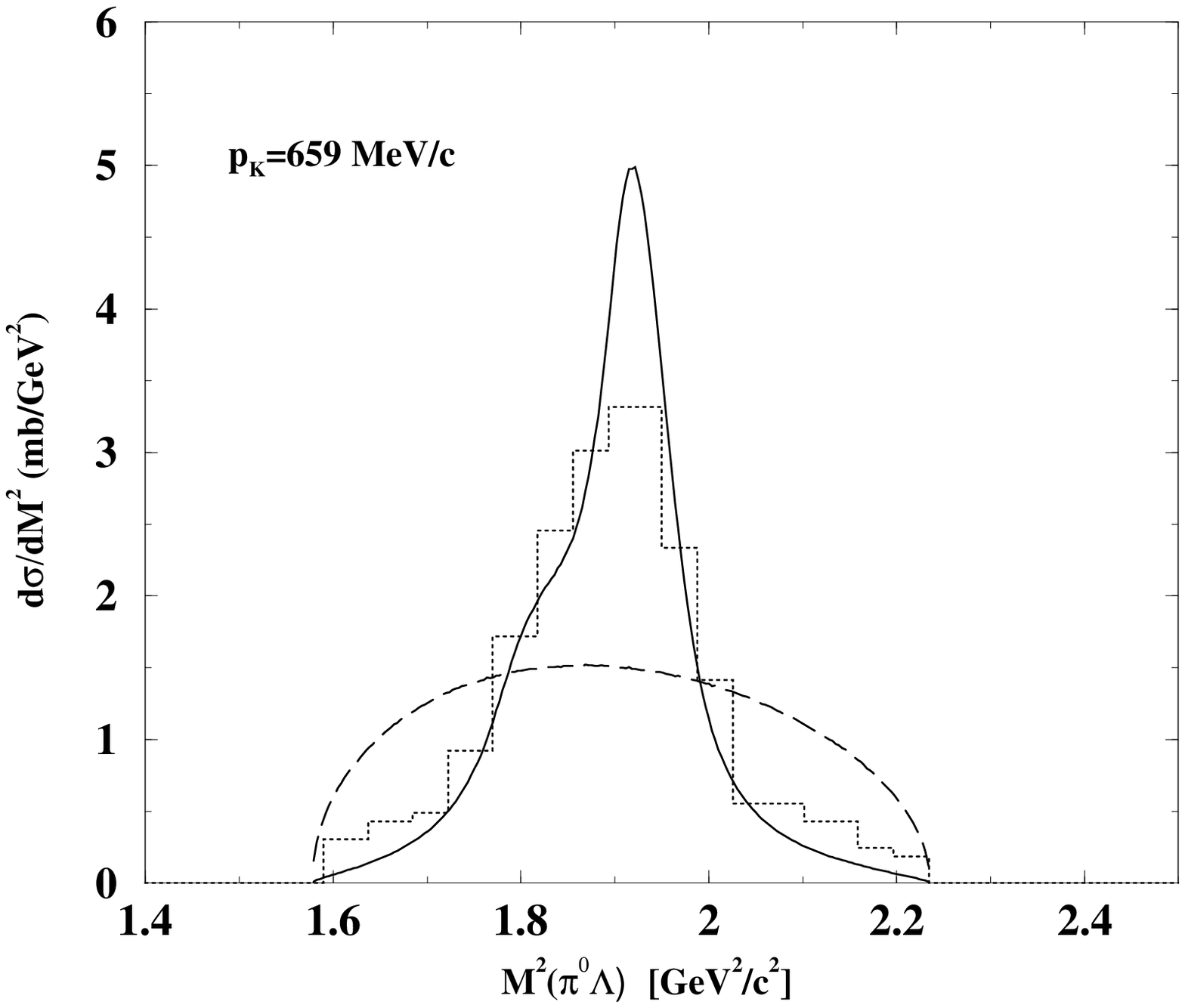}}
\epsfxsize=8cm
\centerline{\epsfbox{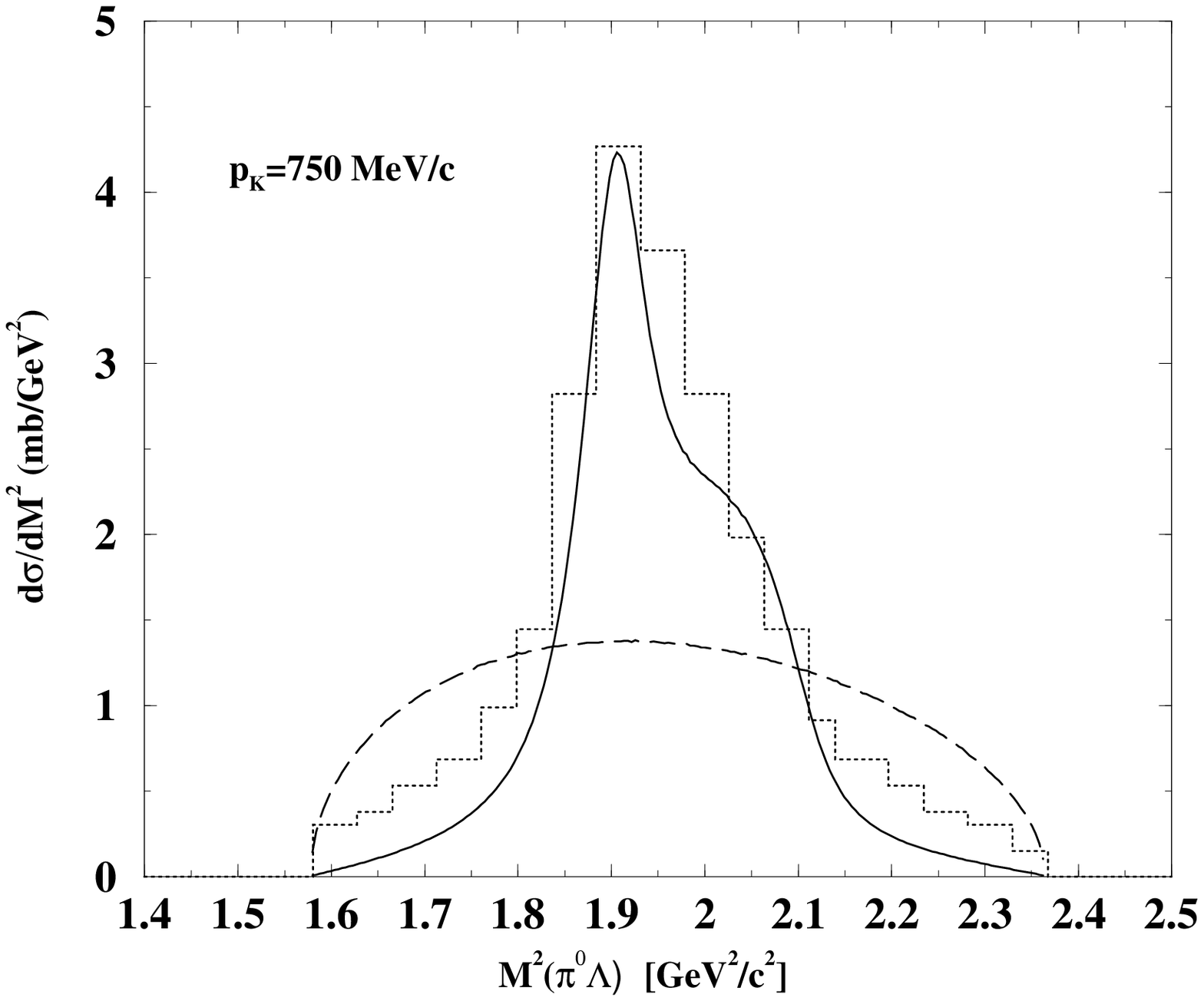}}
\caption{$d\sigma/dM^2$  as a function
of the invariant mass of $\pi^0\Lambda$  for two values of the $K^-$ momentum in CM;
 Left: 659 MeV
and Right:750 MeV. Solid lines represent our results. The dotted histograms are
the experimental results from~\protect\cite{Prakhov:2004ri} normalized to the total experimental
cross section. The dashed lines indicate the phase space normalized to the
theoretical cross section.}
\label{dsdmfig}
\end{figure}

In fig. \ref{dsdmfig} we compare the predictions for the invariant mass
distribution of $\pi^0 \Lambda$ with experiment. The shape of the $\Sigma(1385)$
is clearly visible, although, as in Section 9, the symmetry of the pions induces
a background from the "wrong" $\pi^0 \Lambda$ couples. The agreement of the
theory and experiment gives a strong support to the idea of the $\Lambda(1520)$
being a quasibound state of $\pi \Sigma(1385)$, but a stronger support would be
provided by the measurements at lower energy showing the predicted excitation of
the $\Lambda(1520)$ resonance. 

\section{ $\bar{K}$ in nuclei}
 Next we address the properties of the $\bar{K}$ in the 
nuclear medium which have been studied in  \cite{knuc}. The work is based on
the elementary $\bar{K} N$ interaction which has been discussed above,
using a coupled channel unitary approach with chiral Lagrangians.

The coupled channel formalism requires to evaluate the transition
amplitudes between the different channels that can be built from
the meson and baryon octets. For $K^- p$ scattering there are 10 such
channels, namely $K^-p$, $\bar{K}^0 n$, $\pi^0
\Lambda$, $\pi^0 \Sigma^0$,
$\pi^+ \Sigma^-$, $\pi^- \Sigma^+$, $\eta \Lambda$, $\eta
\Sigma^0$,
$K^+ \Xi^-$ and $K^0 \Xi^0$. In the case of $K^- n$ scattering
the coupled channels are: $K^-n$, $\pi^0\Sigma^-$,
 $\pi^- \Sigma^0$, $\pi^- \Lambda$, $\eta
\Sigma^-$ and
$K^0 \Xi^-$.

 In order to evaluate
the $\bar{K}$ selfenergy in the medium, one needs first to include the medium 
modifications 
in the $\bar{K} N$ amplitude, $T_{\rm
eff}^{\alpha}$ ($\alpha={\bar K}p,{\bar K}n$), and then perform the
integral over the nucleons in the Fermi sea: 

\begin{equation}
\Pi^s_{\bar{K}}(q^0,{\vec q},\rho)=2\int \frac{d^3p}{(2\pi)^3}
n(\vec{p}) \left[ T_{\rm eff}^{\bar{K}
p}(P^0,\vec{P},\rho) +
T_{\rm eff}^{\bar{K} n}(P^0,\vec{P},\rho) \right] \ ,
\label{eq:selfka}
\end{equation}

The values
$(q^0,\vec{q}\,)$ stand now for the energy and momentum of the
$\bar{K}$ in the lab frame, $P^0=q^0+\varepsilon_N(\vec{p}\,)$,
$\vec{P}=\vec{q}+\vec{p}$ and $\rho$ is the nuclear matter density.

We also include a p-wave contribution to the ${\bar K}$ 
self-energy coming from the coupling of the ${\bar K}$ meson to
hyperon-nucleon hole ($YN^{-1}$) excitations,
with $Y=\Lambda,\Sigma,\Sigma^*(1385)$. The vertices $MBB^\prime$ are 
easily derived from
the $D$ and $F$ terms of Eq.~(1). The explicit expressions can be seen in
 \cite{knuc}. 
 
 At this point it is interesting to recall three different approaches to the
 question of the $\bar{K}$ selfenergy in the nuclear medium. The first
 interesting realization was the one in \cite{koch94,wkw96,waas97}, 
 where the Pauli blocking in the intermediate nucleon states 
 induced a shift of the $\Lambda(1405)$ resonance to higher
 energies and a subsequent attractive $\bar{K}$ selfenergy. The work of
 \cite{lutz} introduced a novel an interesting aspect, the selfconsistency.
 Pauli blocking required a higher energy to produce the resonance, but having a
 smaller kaon mass led to an opposite effect, and as a consequence the
 position of the resonance was brought back to the free position. Yet, a 
 moderate
 attraction on the kaons still resulted, but weaker than anticipated from the
 former work.  The work of \cite{knuc} introduces some novelties. It
 incorporates the selfconsistent treatment of the kaons done in \cite{lutz} 
 and in addition it also includes the selfenergy of the pions, which are let to
 excite ph and $\Delta h$ components. It also includes the  mean field
 potentials of the baryons.  
The obvious consequence of the work of \cite{knuc}
is that the spectral function of the kaons  gets much wider than in the two
former approaches because one is including new decay channels for the
$\bar{K}$ in nuclei.

  In the work of \cite{zaki} the kaon selfenergy discussed above has been 
  used for the case of kaonic atoms, where there are
abundant data to test the theoretical predictions. One uses the Klein 
Gordon equation and obtains two families of states. One
family corresponds to the atomic states, some of which are those already  
measured, and 
which have  energies around or below 1 MeV and widths
of about a few hundred KeV or smaller. The other family corresponds to 
states which are nuclear deeply bound states, with
energies of 10 or more MeV and widths around 100 MeV. 

With the $K^-$ many body decay channels included in our approach, the resulting
widths of the deeply bound $K^-$ states (never bound by more than 50 MeV) are
very large (of the order of 100 MeV) and, hence, there is no room for narrow
deeply bound $K^-$ states which appear in some oversimplified theoretical
approaches.  A recent phenomenological work \cite{Mares:2004pf} considering the
$K^- NN \to \Lambda N \Sigma N$ nuclear kaon absorption channels, which are also
incorporated in \cite{knuc}, also reaches the conclusion that in the unlikely
case that there would be deeply bound kaonic atoms they should have necessarily
a large width.

\section{ $\phi$ decay in nuclei}

Let us say a few words about the $\phi$ decay in nuclei. The work 
reported here \cite{phi} follows closely the lines of
\cite{klingl,norbert}, however, it uses the updated $\bar{K}$ 
selfenergies of \cite{knuc}.  In the present
case the $\phi$ decays primarily in $K\bar{K}$, but these kaons can 
now interact with the medium as discussed previously.    For
the selfenergy of the $K$, since the $KN$ interaction is not too strong 
and there are no resonances, the $t\rho$ approximation is
sufficient.
\begin{figure}[htb]
\epsfxsize=8cm
\centerline{\epsfbox{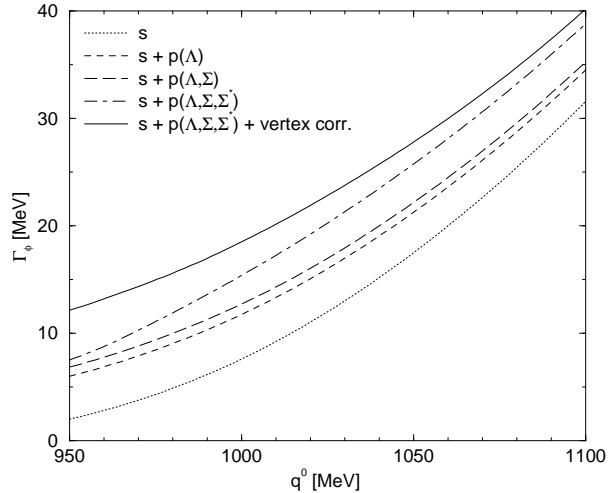}}
 \caption{$\phi$ width at $\rho=\rho_0$.}
\label{phifig}
\end{figure}
In fig. \ref{phifig} we show the results for the $\phi$ width at $\rho=\rho_0$ 
as a function of the mass of the $\phi$, separating the
contribution from the different channels. What we observe is that the 
consideration of the s-wave $\bar{K}$-selfenergy is
responsible for a sizable increase of the width in the medium, but 
the p-wave is also relevant, particularly the $\Lambda h$
excitation and the $\Sigma^*h$ excitation. It is also interesting to 
note that the vertex corrections \cite{beng}  (Yh loops attached
to the $\phi$ decay vertex)  are now present and do not cancel off 
shell contributions like in the case of the scalar mesons. Their
contribution is also shown in the figure and has about the same 
strength as the other p-wave contributions.  The total width of
the $\phi$ that we obtain is about 22 MeV at $\rho=\rho_0$, about 
a factor two smaller than the one obtained in
\cite{klingl,norbert}. A recent evaluation of the $\phi$ selfenergy in the
medium \cite{dani} similar to that of \cite{phi}, in which also the real part is evaluated,
leads to a width about 20 percent larger than that of \cite{phi}. The important 
message from all these works is, however,  the 
nearly one order of magnitude increase of the width with respect to
the free one.

\section{$\eta$ selfenergy and eta bound in nuclei}
 The method of \cite{knuc} has also been used recently to determine the $\eta$ selfenergy
   in the nuclear medium \cite{Inoue:2002xw}.  One obtains a potential at threshold of the order 
   of (-54 -i29) MeV at
   normal nuclear matter, but it also has a strong energy dependence due to the proximity of
   the $N^*(1535)$ resonance and its appreciable modification in the nuclear
   medium. 
   
 To compute the $\eta-$nucleus bound states, we solve the Klein-Gordon 
 equation (KGE) with the $\eta-$selfenergy, 
 $\Pi_\eta(k^0,r) \equiv \Pi_\eta(k^0,\vec 0,\rho(r))$, obtained using the local
 density approximation.
 We have then:
 \begin{equation}
   \left[ -\vec\nabla^2 + \mu^2 +  \Pi_\eta(\mbox{Re}[E],r) \right] \Psi 
  = E^2 \Psi
 \end{equation}
 where $\mu$ is the $\eta-$nucleus reduced mass, the real part of $E$  
 is the total meson energy, including its mass, and the imaginary part
 of $E$, with opposite sign, is the half-width $\Gamma/2$ of the state. 
 The binding energy $B<0$ is defined as $B = \mbox{Re}[E]-m_{\eta}$.

 The results  from \cite{Garcia-Recio:2002cu} are shown in Table 2  for the 
 energy dependent potential.  On the
 other hand we see that the half widths of the states are  
 large, larger in fact
 than the binding energies or the separation energies between neighboring 
 states. 
 
\begin{table}[t]
\begin{center}
\caption{ (B,$-\Gamma/2$) for $\eta-$nucleus bound states 
calculated with the energy dependent potential. Units in MeV}
\label{tbl:statedep}
\footnotesize
\begin{tabular}{c|cccccc}
\hline
\hline 
  &   $^{12}$C  & $^{24}$Mg    &  $^{27}$Al    &  $^{28}$Si    &   $^{40}$Ca   &  $^{208}$Pb    \\
\hline 
1s&($-$9.71,$-$17.5)&($-$12.57,$-$16.7)&($-$16.65,$-$17.98)&($-$16.78,$-$17.93)&($-$17.88,$-$17.19)&($-$21.25,$-$15.88) \\
1p&             &              &( $-$2.90,$-$20.47)&( $-$3.32,$-$20.35)&( $-$7.04,$-$19.30)&($-$17.19,$-$16.58) \\
1d&             &              &               &               &               &($-$12.29,$-$17.74) \\
2s&             &              &               &               &
&($-$10.43,$-$17.99) \\
1f&             &              &               &               &               &( $-$6.64,$-$19.59) \\
2p&             &              &               &               &               &( $-$3.79,$-$19.99) \\
1g&             &              &               &               &               &( $-$0.33,$-$22.45) \\
\hline
\hline
\end{tabular}
\normalsize
\end{center}
\end{table}

   With the results obtained here it looks like the chances to see distinct
 peaks corresponding to $\eta$ bound states are not too big. 
 
   On the other hand one can look at the results with a more optimistic view
 if one simply takes into account that experiments searching for these states
 might not see them as peaks, but they should see some clear
 strength below threshold in the $\eta$ production experiments.
 The range by which this strength would go into the bound region
 would measure the combination of half width and binding energy. 
 Even if this is less information than the values of the energy and
 width of the states, it is by all means a relevant information to gain some
 knowledge on the $\eta$ nucleus optical potential.  
 
\section{Experiments to determine the $\phi$ width in the medium}

Recently there has been an experiment \cite{ishi} designed to determine the
width of the $\phi$ in nuclei.  Unlike other experiments proposed which aim at
determining the width from the $K^+ K^-$ invariant mass distribution and which
look extremely difficult \cite{Oset:2000na,muhlich}, the experiment done in
Spring8/Osaka uses a different philosophy, since it looks at the A dependence of
the $\phi$ photoproduction cross section. The idea is that the $\phi$ gets absorbed
in the medium with a probability per unit length equal to 
\begin{equation}
-Im \Pi /q
\end{equation}
where $\Pi$ is the $\phi$ selfenergy in the medium ($\Gamma=-Im \Pi /
\omega_{\phi}$). The bigger the nucleus the more $\phi$ get absorbed and there
is a net diversion from the A proportionality expected from a photonuclear
reaction.
 The method works and one obtains a $\phi$ width in the medium which is given in
 \cite{ishi} in terms of a modified $\phi N$ cross  section in the nucleus
 sizably larger than the free one.  Prior to these experimental results there is
 a theoretical calculation in \cite{luis} adapted to the set up of the
 experiment of \cite{ishi}, based on the results for the $\phi $ selfenergy
 reported above.  The results agree only qualitatively with the experimental
 ones, the latter ones indicating that the $\phi$ selfenergy in the medium could
 be even  larger than the calculated one. However, although it has been fairly
 taken into account in the experimental analysis, there is an inconvenient in the
 reaction of \cite{ishi}, since there is a certain contamination of coherent
 $\phi$ production which blurs the interpretation of the data.
 
\begin{figure}[htbp]
\epsfxsize=11cm
\centerline{\epsfbox{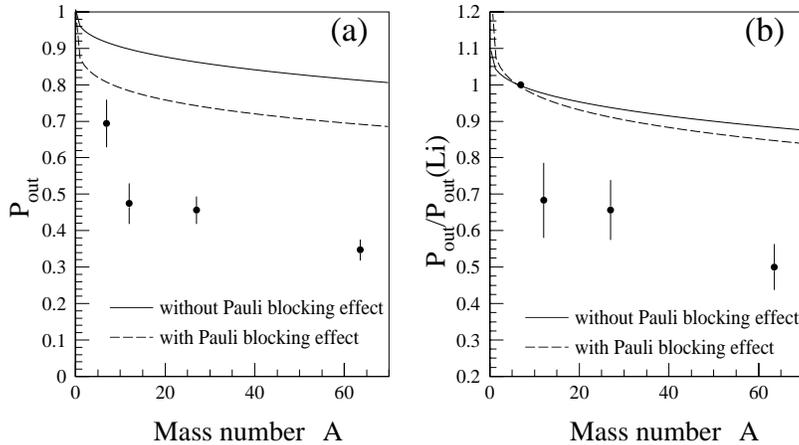}}
\caption{
(a)~The probability $P_{\rm out}\!=\!\sigma_A/(A\sigma_N)$.
The overall normalization error~(18\%)
is not included. 
The solid and dashed curves show the theoretical calculations
given by Cabrera et al.~\protect\cite{luis} without and with
Pauli-blocking correction for the $\phi$ meson scattering angle
in the laboratory frame of 0${}^\circ$, respectively.
(b)~The ratio $P_{\rm out}/P_{\rm out}{\rm (Li)}$. 
The solid and dashed curves show the theoretical calculations same as (a).}
\label{fig:ishi}
\end{figure}

 This can be seen in fig. \ref{fig:ishi} from \cite{ishi}. In the figure to the
 right, the cross sections are normalized to the one of $^7 Li$
 
There is an apparent discrepancy of the experimental results with the theory. 
Yet, one could assume that there is much contamination of coherent $\phi$
production in $^7 Li$, which certainly contains the largest amount since
coherent production is killed by the nuclear form factors, much smaller in heavy
nuclei for the momentum transfers involved in the reaction. If one removes the 
$^7Li$ datum and normalizes all the cross sections to $^{12} C$, then the
agreement between theory and experiment is much better.
 
\begin{figure}[htb]
\epsfxsize=9cm
\centerline{\epsfbox{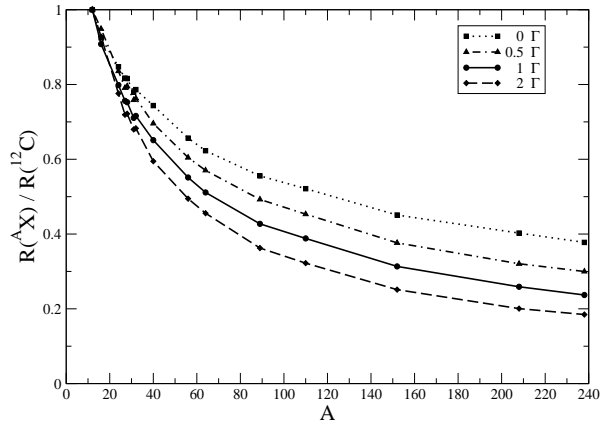}}
\caption{Ratio of the nuclear cross section normalized to
$^{12}C$ for $T_p=2.83\textrm{ GeV}$ multiplying the $\phi$ width
in the medium, $\Gamma$ around 27 MeV, by different factors.} 
\label{phiprod}
\end{figure}

   In order to use the same idea of the A dependence and get rid of the coherent
$\phi$ production, a new reaction has been suggested in \cite{magas} which could
be implemented in  a facility like COSY. The idea is to measure the $\phi$
production cross section in different nuclei through the reaction
\begin{equation}
p A \to  \phi X
\end{equation}
The calculation is done assuming one step production, and two step production
with a nucleon or $\Delta$ in the intermediate states,  allowing for the loss of
$\phi$ flux as the $\phi$ is absorbed in its way out of the nucleus. Predictions
for the cross sections normalized to the one of $^{12}C$ are shown in 
fig. \ref{phiprod}
where it is shown that with a precision of 10 percent in the experimental 
ratios one could
disentangle between the different curves in the figure and easily determine if
the width is one time, two times etc, the width determined in ref.
\cite{phi,dani} which is about 27 MeV for normal nuclear matter density. An
experiment of this type can be easily performed in the COSY facility, where
hopefully it will be done in the near future.

\section{Conclusions}
The use of chiral Lagrangians for the meson baryon interaction and the unitary
extensions of chiral perturbation theory have allowed to face a large amount of
problems which were barred to standard perturbation techniques. It has
opened the door to the study of many baryonic resonances which qualify neatly as
dynamically generated resonances, or quasibound states of meson baryon. Thanks
to this, a quantitative description of the meson baryon interaction at 
intermediate energies is now possible and with this, an important systematics has
been introduced in the many body problem to face issues on the renormalization of
hadron properties in the nuclear medium. The constructed scheme is rather powerful
and allows to make predictions which consecutive experiments are proving right.
Two of these predictions, the two states for the $\Lambda(1405)$, and the nature
of the $\Lambda(1520)$ as a quasibound state of $\pi \Sigma(1385)$, have found
strong support from two very recent experiments.  Ongoing experiments on the
$\phi$ width in the nuclear medium should soon provide reliable information on
this so long sought important magnitude. Altogether, one is seeing through all
this work that chiral dynamics is a key ingredient that allows a unified
description of much of the hadronic world at low and intermediate energies.
 
\section*{Acknowledgments}
D.C. and  L.R. acknowledge support from the 
Ministerio de Educaci\'on y Ciencia.
This work is partly supported by the Spanish CSIC and JSPS collaboration, the
 DGICYT contract number BFM2003-00856,
and the E.U. EURIDICE network contract no. HPRN-CT-2002-00311. 
This research is part of the EU
      Integrated Infrastructure Initiative
      Hadron Physics Project under contract number
      RII3-CT-2004-506078.

\vfill\eject
\end{document}